\documentclass[pdflatex,sn-mathphys-num]{sn-jnl}
\usepackage[utf8]{inputenc}
\usepackage{microtype}
\usepackage{amsmath,amssymb,amsthm}
\usepackage{xspace}
\usepackage{hyperref}
\usepackage{cleveref}
\usepackage{anslistings}
\usepackage{colonequals}
\usepackage{subcaption}

\renewcommand{\orcidlogo}{\includegraphics[height=10pt]{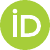}}
\renewcommand{\orcid}[1]{\href{https://orcid.org/#1}{\orcidlogo}}

\newcommand{\dune}{\textsc{dune}\xspace}
\newcommand{\dunecommon}{\textsc{dune-common}\xspace}
\newcommand{\dunegeometry}{\textsc{dune-geometry}\xspace}
\newcommand{\dunegrid}{\textsc{dune-grid}\xspace}
\newcommand{\duneistl}{\textsc{dune-istl}\xspace}
\newcommand{\dunelocalfunctions}{\textsc{dune-localfunctions}\xspace}
\newcommand{\dunefunctions}{\textsc{dune-functions}\xspace}
\newcommand{\dunetypetree}{\textsc{dune-typetree}\xspace}
\newcommand{\dunealugrid}{\textsc{dune-alugrid}\xspace}

\newcommand{\dunefem}{\textsc{dune-fem}\xspace}
\newcommand{\dunepdelab}{\textsc{dune-pdelab}\xspace}
\newcommand{\dunefufem}{\textsc{dune-fufem}\xspace}
\newcommand{\dunecurvedgrid}{\textsc{dune-curvedgrid}\xspace}
\newcommand{\dunecurvedgeometry}{\textsc{dune-curvedgeometry}\xspace}
\newcommand{\dunecopasi}{\textsc{dune-copasi}\xspace}

\newcommand{\dunevtk}{\textsc{dune-vtk}\xspace}
\newcommand{\amdis}{AMDiS\xspace}
\newcommand{\dumux}{DuMu\textsuperscript{X}\xspace}
\newcommand{\kaskade}{Kaskade~7\xspace}
\newcommand{\opm}{OPM\xspace}
\newcommand{\dunecurvilineargeometry}{\textsc{dune-curvilineargeometry}\xspace}

\begin{document}

\title[Dune 2.10]{The Distributed and Unified Numerics Environment (DUNE), Version 2.10}

\author[1]{\fnm{Markus} \sur{Blatt} \orcid{0009-0003-1869-2834}}\affil[1]{Dr. Markus Blatt --- HPC-Simulation-Software \& Services, Heidelberg}

\author[2]{\fnm{Samuel} \sur{Burbulla} \orcid{0000-0002-2566-9777}}\affil[2]{\orgdiv{Institut für Angewandte Analysis und Numerische Simulation}, \orgname{Universität Stuttgart}, \orgaddress{\street{ Pfaffenwaldring 57}, \postcode{70569} \city{Stuttgart}, \country{Germany}}}

\author[3]{\fnm{Ansgar} \sur{Burchardt}}\affil[3]{\orgdiv{Zentrum für Informationsdienste und Hochleistungsrechnen}, \orgname{Technische Universität Dresden}, \orgaddress{\postcode{01062} \city{Dresden}, \country{Germany}}}

\author[4]{\fnm{Andreas} \sur{Dedner}}\affil[4]{\orgdiv{Warwick Mathematics Institute}, \orgname{University of Warwick}, \orgaddress{\city{Coventry} \postcode{CV4~7AL}, \country{UK}}}

\author[5]{\fnm{Christian} \sur{Engwer} \orcid{0000-0002-6041-8228}}\affil[5]{\orgdiv{Institut für Numerische und Angewandte Mathematik}, \orgname{Universität Münster}, \orgaddress{\street{Einsteinstr. 62}, \postcode{48149} \city{Münster}, \country{Germany}}}

\author[6]{\fnm{Carsten} \sur{Gräser} \orcid{0000-0003-4855-8655}}\affil[6]{\orgdiv{Department Mathematik}, \orgname{Friedrich-Alexander-Universität Erlangen-Nürnberg}, \orgaddress{\street{Cauerstraße 11}, \postcode{91058} \city{Erlangen}, \country{Germany}}}

\author[7]{\fnm{Christoph} \sur{Grüninger}}\affil[7]{Unaffiliated, \city{Stuttgart}, \country{Germany}}

\author[8]{\fnm{Robert} \sur{Klöfkorn} \orcid{0000-0001-9664-0333}}\affil[8]{\orgdiv{Center for Mathematical Sciences}, \orgname{Lund University}, \orgaddress{P.O. Box 117, \postcode{221~00}, \city{Lund}, \country{Sweden}}}

\author[9]{\fnm{Timo} \sur{Koch} \orcid{0000-0003-4776-5222}}\affil[9]{\orgdiv{Department of Mathematics}, \orgname{University of Oslo}, \orgaddress{\postcode{0316} \city{Oslo}, \country{Norway}}}

\author[10,1]{\fnm{Santiago} \sur{Ospina} \sfx{De Los Ríos} \orcid{0000-0003-0814-9670}}\affil[10]{\orgdiv{Interdisciplinary Center for Scientific Computing}, \orgname{Heidelberg University}, \orgaddress{\street{Im Neuenheimer Feld 205}, \postcode{69120} \city{Heidelberg}, \country{Germany}}}

\author*[11]{\fnm{Simon} \sur{Praetorius} \orcid{0000-0002-1372-4708}}\email{simon.praetorius@tu-dresden.de}
\affil*[11]{\orgdiv{Institut für Wissenschaftliches Rechnen}, \orgname{Technische Universität Dresden}, \orgaddress{\postcode{01062} \city{Dresden}, \country{Germany}}}

\author[12]{\fnm{Oliver} \sur{Sander} \orcid{0000-0003-1093-6374}}\affil[12]{\orgdiv{Institut für Numerische Mathematik}, \orgname{Technische Universität Dresden}, \orgaddress{\postcode{01062} \city{Dresden}, \country{Germany}}}

\abstract{Version 2.10 of the Distributed and Unified Numerics Environment (\dune) introduces a range of enhancements across its core and extension modules, with a continued emphasis on modern C++ integration and improved usability. This release extends support for C++20 features, particularly concepts, through comprehensive refinements in {\dunecommon} and {\dunegrid}, enabling safer and more expressive generic programming paradigms. A notable advancement is the improved support for curved geometries, including new geometry implementations and a more flexible interface. Data structures have been modernized through native support for \texttt{std::mdspan} and \texttt{std::mdarray}, performance improvements in sparse matrices, and tools for visualization of matrix patterns. The build system has been restructured towards a modern CMake workflow, emphasizing target-based configuration and improved automation. Furthermore, new local finite elements have been introduced to broaden numerical capabilities. The release also brings updates across {\dune} extensions, as well as improvements to infrastructure and module-level components.}

\keywords{numerical software; finite elements; finite volumes; grid-based numerical analysis; DUNE.}

\maketitle

\section{Introduction}

The Distributed and Unified Numerics Environment ({\dune})~\cite{dunepaperI:08,dunepaperII:08,BlattBastian2007DuneIstl,BlattBastian2008Generic} is a modular software framework for the numerical solution of partial differential equations (PDEs) based on grid-based discretization methods. It provides a flexible and efficient platform for implementing a wide range of numerical schemes, including finite element, finite volume, and discontinuous Galerkin methods. The {\dune} ecosystem is structured into a set of core modules responsible for fundamental functionality such as grid management, geometry representation, linear algebra, and local finite element implementations, complemented by a growing number of extension and application modules.

Building on this modular structure, each {\dune} module provides interoperable data structures and algorithmic building blocks. The {\dune} core and extension modules --- which constitute the focus of this paper --- are not intended to provide comprehensive, high-level application interfaces for grid-based PDE simulations from an end-user perspective. Rather, they serve as low-level infrastructure for constructing such applications. For simulation frameworks that integrate the various {\dune} components into complete numerical applications, we refer the interested reader to several {\dune}-based projects, including {\amdis}~\cite{Praetorius2025AMDiS}, {\dumux}~\cite{KochEtAl2021Dumux}, {\dunefem}~\cite{DednerEtAl2010DuneFem}, {\dunefufem}~\cite{dune-fufem}, {\dunepdelab}~\cite{BlattHeimannMarnach2010Generic}, {\kaskade}~\cite{Gtschel2021Kaskade}, and {\opm}~\cite{Rasmussen2021}.

A comprehensive overview of the framework and its development up to version~2.7 was presented in~\cite{BastianEtAl2021Dune}. Since then, the {\dune} project has undergone substantial further development, particularly in the areas of modern C++ adoption, performance optimization, and support for advanced geometry and build system features.

This paper presents the key improvements and interface changes introduced in version~2.10 of the {\dune} core modules --- namely {\dunecommon}, {\dunegeometry}, {\dunegrid}, {\duneistl}, and {\dunelocalfunctions}~--- released on September~4, 2024. In addition, we highlight noteworthy developments in several associated extension modules that are part of the broader {\dune} ecosystem.

\subsection{Code availability}
The code of the {\dune} core modules and extension modules is available in their corresponding repositories in \href{https://gitlab.dune-project.org/core}{gitlab.dune-project.org/core}, \href{https://gitlab.dune-project.org/staging}{gitlab.dune-project.org/staging}, and \href{https://gitlab.dune-project.org/extensions}{gitlab.dune-project.org/extensions} in the Git release branch \texttt{releases/2.10}. The software is licensed under version 2 of the GNU General Public License, with a special exception for linking and compiling against {\dune}; more details can be found at \href{https://dune-project.org/about/license}{dune-project.org/about/license}.

The code examples provided throughout the paper focus on illustrating techniques, algorithms, and data structures, omitting standard library includes and the \texttt{main()} function for readability. Full working code examples are available in a supplementary document of the paper referenced by the filenames in the top of the code listings.

\subsection{Development and community}
The {\dune} ecosystem is a continuously evolving open-source project, actively maintained and developed by a community of contributors from both academia and industry. For up-to-date information on recent developments, ongoing discussions, contact details, and opportunities to contribute, we refer interested readers to the official {\dune} website at \href{https://www.dune-project.org}{dune-project.org}, as well as the {\dune} GitLab instance at \href{https://gitlab.dune-project.org/}{gitlab.dune-project.org}, which hosts the aforementioned projects, issue trackers, and additional resources.

\subsection{Outline of the paper}
This paper starts by introducing the new features added to the {\dune} core modules in \cref{sec:new-core-feature}, including changes to the build system (\cref{sec:buildsystem}). Changes outside the core modules, such as in extension modules, are discussed in \cref{sec:development-in-extensions}. The paper is concluded by a summary list of noteworthy changes in all discussed modules in \cref{sec:further-noteworthy-changes}.

\section{New core features}\label{sec:new-core-feature}

The 2.10 release of the {\dune} core modules introduces a range of incremental yet impactful improvements. These include continued code modernization, improved maintainability, enhanced developer tooling, and minor feature additions across several core components. While the release does not focus on major interface overhauls, it lays important groundwork for future capabilities, particularly through greater adoption of C++20 language features and better support for modern development workflows. Notably, Python bindings are now enabled by default across supported modules, further improving accessibility for users and facilitating integration into Python-based workflows.

\subsection{Growing support for C++ concepts}

\subsubsection{C++ concepts in dune-common}

The C++20 standard introduces \emph{concepts} and type constraints as a formal mechanism for constraining template parameters, offering improved clarity and compile-time error diagnostics compared to traditional SFINAE-based techniques\footnote{SFINAE, Substitution failure is not an error, eliminates candidates of a function overload set without raising a compiler error.}. As many {\dune} interfaces rely heavily on SFINAE checks, this release marks the beginning of a gradual migration toward the use of concepts to formalize and document interface requirements.

In {\dunecommon}, a foundational set of concepts is now provided under the \texttt{Dune::Concept} namespace. These are currently employed primarily for internal testing and for defining more structured grid-related concepts, as further detailed in \cref{sec:dunegrid-concepts}.
\renewcommand*{\thecodefile}{concepts.cc}
\begin{cppcode}
#include <dune/common/concepts.hh>

static_assert(Dune::Concept::RandomAccessContainer< Dune::ReservedVector<double,3> >);
static_assert(Dune::Concept::Hashable< Dune::bigunsignedint<128> >);
\end{cppcode}
Since concepts require the C++20 standard to be enabled, which is not done by default\footnote{A new C++ standard version can be set by providing the CMake parameter \texttt{-DCMAKE\_CXX\_STANDARD=20} during configuration of {\dune} modules.} in {\dune} 2.10, the concept definitions and tests are only available if supported and the preprocessor constant \texttt{DUNE\_ENABLE\_CONCEPTS} is set. This constant can also be used to conditionally enable tests or some code to use concepts when available.

\subsubsection{dune-grid concepts}\label{sec:dunegrid-concepts}
Starting with {\dune} 2.10, we have also released an experimental version of C++20 concepts for the {\dune} grid interface. When enabled, concepts for the grid components are available through the \texttt{dune/grid/concepts.hh} header and can be used to constrain types throughout the code:

\renewcommand*{\thecodefile}{gridconcepts.cc}
\begin{cppcode}
#include <dune/grid/concepts.hh>

// constrain type 'Grid' to be a dune grid
template <Dune::Concept::Grid Grid>
void integrate (const Grid& grid, /*...*/) {
  // constrain type of 'gv' to be a dune grid view
  Dune::Concept::GridView auto gv = grid.leafGridView();

  // constrain type of 'is' to be a dune grid index set
  const Dune::Concept::IndexSet auto& is = gv.indexSet();

  // constrain type of 'e' to be a dune grid entity
  for (const Dune::Concept::Entity auto& e : elements(gv)) {
    // constrain type of 'geo' to be a dune grid geometry
    Dune::Concept::Geometry auto geo = e.geometry();
    // constrain type of 'i' to be a dune grid intersection
    for (const Dune::Concept::Intersection auto& i : intersections(gv,e))
      // ...
  }
}
\end{cppcode}

Note that most of the concept checks in the above snippet are superfluous and were written to demonstrate usage. Constraining a type to be a \texttt{Dune::\allowbreak Concept::\allowbreak Grid} already guarantees that the method \texttt{.leafGridView()} will return a type that satisfies the \texttt{Dune::\allowbreak Concept::\allowbreak GridView} concept. A type that is a \texttt{Dune::\allowbreak Concept::\allowbreak Entity} also guarantees that \texttt{.geometry()} returns a type that satisfies the \texttt{Dune::\allowbreak Concept::\allowbreak Geometry} concept, and so on\footnote{The exception is the method \texttt{.grid()} in the \texttt{Dune::\allowbreak Concept::\allowbreak GridView} due to its recursive dependency on the grid concept. Only a reduced concept will be checked instead.}.

Having these concepts guarantees that C++ types have the properties and functions expected in a dune grid. If the incoming type does not fulfill such concept constraints, the compiler will immediately issue a compile-time error explaining the situation. This prevents the use of incomplete dune grid types in contexts where the grid interface is expected.

An advantage of these concepts is that it becomes easier to develop grids that satisfy the {\dunegrid} interface. In previous releases, to check if a grid compiles with the {\dunegrid} interface, the grid type must already contain the entire interface. With the new grid concepts, the individual interfaces of the grid (\texttt{Dune::\allowbreak Concept::\allowbreak GridView}, \texttt{Dune::\allowbreak Concept::\allowbreak IdSet}, \texttt{Dune::\allowbreak Concept::\allowbreak Intersection}, etc.) can be checked independently and later used to implement the class that satisfies the full \texttt{Dune::\allowbreak Concept::\allowbreak Grid} concept.
 \subsection{New tools for meta-programming}

Many components in the {\dune} framework --- including grids, linear solvers, and function space hierarchies --- rely extensively on compile-time (template) meta-programming. To support this, the {\dunecommon} module provides a collection of utilities that simplify and standardize such techniques across the framework. With each release, these utilities are extended and refined to align with ongoing developments in the C++ standard library, aiming to improve code readability, expressiveness, and maintainability.

\subsubsection{Utilities for \texttt{std::integral\_constant} and \texttt{std::index\_sequence}}

Compile-time index values are commonly used across several {\dune} modules. For example, {\duneistl} employs them to index heterogeneous data structures such as \texttt{Dune::MultiTypeBlockVector}; {\dunegrid} uses them to implement tests for all provided static codimensions of a grid; and {\dunetypetree} applies them to navigate node positions in type trees. To facilitate such use cases, {\dunecommon} now introduces a suite of utilities for generating and manipulating integral constants.

Predefined compile-time indices are available since early versions of {\dune} in the \texttt{Dune::Indices} namespace, offering a convenient range from \texttt{\_0} to \texttt{\_19}. Additionally, user-defined indices can now be created directly using the \texttt{""\_ic} user-defined literal, enabling the construction of integral constants for arbitrary values. These indices support a range of operations, including the definition and traversal of compile-time index ranges:
\renewcommand*{\thecodefile}{indexrange.cc}
\begin{cppcode}
#include <dune/common/hybridutilities.hh>
#include <dune/common/rangeutilities.hh>

using namespace Dune::Indices;
auto static_range = Dune::range(0_ic, 10_ic); // range-like container [0_ic, 1_ic,..., 9_ic]
auto dyn_range = Dune::range(0, 10); // range of (dynamic) indices [0, 1,..., 9]

// loop over the static range and get a static index in each iteration
Dune::Hybrid::forEach(static_range, [](auto i) {...});

// loop over the dynamic range and get a runtime index in each iteration
Dune::Hybrid::forEach(dyn_range, [](auto i) {...});
\end{cppcode}

The \texttt{Hybrid::forEach} loop used in the example above was introduced in {\dune} version 2.4 to allow flexible implementation of algorithms for data structures that require static or dynamic indexing. This goes even further for data structures that require both dynamic and static indexes, depending on where in the data structure you are accessing an element. An example of such a data structure is a nested \texttt{MultiTypeBlockVector}. We need utilities for transforming and comparing these generalized integers. These are introduced as hybrid function object types, which map a variadic number of (static or dynamic) indices to a new index that is static only if all arguments are static indices, otherwise a dynamic value.
\renewcommand*{\thecodefile}{hybridfunctor.cc}
\begin{cppcode}
#include <dune/common/indices.hh>
#include <dune/common/hybridutilities.hh>

// hybrid functionals
using namespace Dune::Indices;
auto dyn3 = Dune::Hybrid::plus(1_ic, 2);
auto static4 = Dune::Hybrid::minus(6_ic, 2_ic);

// hybrid comparison
assert(not Dune::Hybrid::equal_to(dyn3, static4));
static_assert(Dune::Hybrid::equal_to(static4, 4_ic));
\end{cppcode}

When multiple static indices come into play or need to be expanded, the \texttt{std::integer\_sequence} type is a useful tool. We have extended {\dune} to provide a range-like data structure as well as utilities to access and transform such sequences. The list of functions matches typical container operations, e.g, \texttt{get(seq,index)}, \texttt{front(seq)}, \texttt{back(seq)}, \texttt{filter(predicate,seq)}, \texttt{head(seq)}, and \texttt{tail(seq)} for extracting elements or subsequences, \texttt{push\_front(seq,index)} and \texttt{push\_back(seq,index)} to create a new sequence, \texttt{equal(seq1,seq2)}, \texttt{size(seq)}, \texttt{empty(seq)} and \texttt{contains(seq,index)} to extract information about sequences.
\renewcommand*{\thecodefile}{integersequence.cc}
\begin{cppcode}
#include <dune/common/integersequence.hh>

auto seq = static_range.to_integer_sequence(); // std::index_sequence<0,1,...,9>

// unpack the indices as integral_constants into a function for variadic
// or fold operations
static_assert(Dune::unpackIntegerSequence([](auto... i) { return (i + ...); }, seq) == 45);

// some container functionality
static_assert(Dune::get(seq, 2_ic) == 2); // or get<2>(seq)
static_assert(Dune::front(seq) == 0);
static_assert(Dune::back(seq) == 9);

// compare index sequences
static_assert(not Dune::equal(seq, std::index_sequence<0,2,4,6,8>()));
\end{cppcode}
 \subsection{Support for curved geometries}\label{sec:support-for-curved-geometries}

Accurate representation of complex geometries plays a crucial role in the numerical approximation of PDEs, especially for high-order methods or problems posed on curved or non-flat domains. While existing geometry implementations in {\dune}~--- such as \texttt{Dune::\allowbreak AffineGeometry} and \texttt{Dune::\allowbreak MultiLinearGeometry}\footnote{A \texttt{MultiLinearGeometry} which represents a twisted square embedded in higher-dimensional space, may not be flat.} --- are well-suited for piecewise linear or affine domains, they fall short when it comes to modeling curved or non-linear geometries. To address this, several external modules (e.g., {\dunecurvilineargeometry}~\cite{FominsOswald2016DuneCurvilinearGrid,dune-curvilineargeometry} and {\dunecurvedgeometry}~\cite{PraetoriusStenger2022CurvedGrid,dune-curvedgeometry}) have previously introduced extended functionality.

As of version~2.10, selected capabilities from these extensions have been integrated directly into the {\dunegeometry} module, improving interoperability and reducing external dependencies for users working with curved domains.

\subsubsection{New geometry implementations}\label{sec:new-geometry-implementations}

Version~2.10 introduces the \texttt{Dune::MappedGeometry} class, which generalizes existing geometry interfaces to support mappings from a reference domain into the world (physical) domain via user-defined transformations. Specifically, a \texttt{MappedGeometry} instance combines a reference geometry --- such as the geometry of a reference element or grid element --- with a coordinate transformation function that models a differentiable mapping in the sense of~\cite{EngwerEtAl2017Interface}. That is, the mapping must be differentiable with respect to the coordinates of the reference domain.

This abstraction allows users to represent a wide class of geometries beyond linear approximations, enabling more accurate discretizations in applications where geometric fidelity is critical, such as in higher-order surface finite-element methods or for curved boundary conditions.
\renewcommand*{\thecodefile}{mappedgeometry.cc}
\begin{cppcode}
#include <dune/geometry/mappedgeometry.hh>

// define a differentiable function
struct Surface {
  auto operator() (const Dune::FieldVector<double,2>& x) {
    return Dune::FieldVector<double,3>{x[0], x[1], std::sin(x[0]*x[1])};
  }
  friend auto derivative (const Surface& surface) {
    return [](const Dune::FieldVector<double,2>& x) {
      return Dune::FieldMatrix<double,3,2>{
        {1.0,0.0},
        {0.0,1.0},
        {x[1]*std::cos(x[0]*x[1]),x[0]*std::cos(x[0]* x[1])}};
    };
  }
};
for (const auto& e : elements(gridView)) {
  // construct the mapped geometry
  Dune::MappedGeometry geometry{Surface{}, e.geometry()};
}
\end{cppcode}

Another way to represent parametrized geometries is by providing a mapping in a local function basis. This allows to use existing finite element basis functions and associated coefficients as representation. The differentiation of such a mapping can be done directly in terms of derivatives of the basis functions. A geometry type based on this approach is called \texttt{Dune::LocalFiniteElementGeometry}. Especially in combination with the local finite elements provided in {\dunelocalfunctions} this is a very flexible way to represent discrete geometries.
\renewcommand*{\thecodefile}{localfiniteelementgeometry.cc}
\begin{cppcode}
#include <dune/geometry/localfiniteelementgeometry.hh>
#include <dune/localfunctions/lagrange/lagrangecube.hh>

// define a local finite-element (from dune-localfunction)
using LocalFiniteElement = Dune::LagrangeCubeLocalFiniteElement<double,double,2,order>;
LocalFiniteElement localFE{};

for (const auto& e : elements(gridView)) {
  // projection from local coordinates in the reference element
  // to 2d coordinates in the base grid to 3d global coordinates
  auto X = [geo=e.geometry()](const auto& local) {
    auto x = geo.global(local);
    return Dune::FieldVector<double,3>{x[0], x[1], std::sin(x[0]*x[1])};
  };

  // construct the parametrized geometry, this performs an interpolation
  // of the function X into the local finite element space localFE
  Dune::LocalFiniteElementGeometry geometry{referenceElement(e), localFE, X};
}
\end{cppcode}
In the example, a local function is interpolated into the local finite-element space. An alternative is to provide the local coefficients directly. This allows to represent parametrized geometries and could be used to implement moving/evolving mesh finite-elements.

\subsubsection{Extended geometry interface}\label{sec:extended-geometry-interface}
All geometry types in {\dunegeometry} and also their interface class in {\dunegrid} are extended by the methods \texttt{jacobian()} and \texttt{jacobianInverse()}. Previously, only the transposed version of these methods were available. This was due to the fact that the resulting matrices must be stored in a row-wise fashion and the transposed version made this easier to handle in several cases. However, for vector- and tensor-valued basis functions, such as in Raviart--Thomas or N\'ed\'elec finite elements, the actual Jacobian and its inverse are a more natural choice. For co- and contravariant Piola transforms and their inverse transforms, both the Jacobian and its transpose are sometimes required.
A default implementation for all geometry classes is available, thanks to the recently added utility \texttt{transposedView()} in {\dunecommon}. In geometries with codimension~$> 0$, the inversion of the Jacobian matrix needs to be represented as a pseudo-inverse of that matrix. While the Jacobian transposed allows a right-pseudo-inverse, the non-transposed matrix needs the left-pseudo-inverse representation. The corresponding utility in {\dunegeometry}, \texttt{leftInvA()}, is now aligned to the \texttt{rightInvA()} utility function.
 \subsection{Data structure improvements}

With the evolution of the C++ standard --- most recently in C++23 and upcoming proposals --- the standard library continues to introduce abstractions that improve expressiveness, performance, and memory handling. In {\dune}, many of these developments are directly applicable to core areas such as grid implementations and linear algebra backends. Version~2.10 incorporates several of these modern data structures into the framework, offering forward compatibility with future standard features while maintaining support for C++17.

\subsubsection{Multidimensional arrays with \texttt{mdspan} and \texttt{mdarray}}

The C++23 standard introduces \texttt{std::mdspan}~\cite{mdspan}, a lightweight and non-owning multidimensional view over a contiguous data block, defined by (something like) a pointer, a set of extents, an index layout, and a data access policy. This abstraction enables multi-indexed access with minimal overhead, and is particularly well-suited for matrix and tensor-like operations. A related proposal~\cite{mdarray}, suggests an owning counterpart to \texttt{mdspan}: \texttt{std::mdarray}, which offers a container-like interface for multidimensional data.

To facilitate adoption and experimentation, {\dunecommon} now provides full implementations of both \texttt{mdspan} and \texttt{mdarray} under the namespace \texttt{Dune::Std}. These implementations closely follow the C++ standard and proposal specifications, with only minor deviations to ensure compatibility with the C++17 standard currently required by {\dune} 2.10. These data structures are expected to form the basis of upcoming data structures in future {\dune} releases.
\renewcommand*{\thecodedir}{src/}
\renewcommand*{\thecodefile}{mdspan.cc}
\begin{cppcode}
#include <dune/common/std/mdspan.hh>

std::vector<double> values(10*10*10, 0.0);
Dune::Std::mdspan tensor{values.data(), Dune::Std::extents<std::size_t, 10,10,10>{}};
// or: Dune::Std::mdarray tensor{Dune::Std::extents<std::size_t, 10,10,10>{}, 0.0};
for (std::size_t i0 = 0; i0 < tensor.extent(0); i0++)
  for (std::size_t i1 = 0; i1 < tensor.extent(1); i1++)
    for (std::size_t i2 = 0; i2 < tensor.extent(2); i2++)
      std::cout << tensor(i0,i1,i2); // allows round braces for index access
\end{cppcode}

Tensor data structures based on \texttt{Std::mdarray} are already in development. These include a generalization of \texttt{DenseVector} and \texttt{DenseMatrix} with dynamic dimensions and \texttt{FieldVector} and \texttt{FieldMatrix} for static dimensions. Since we want to add low-level data structures already in {\dunecommon} that may be used by many downstream modules, we have decided not to add a dependency on external libraries, like the kokkos library~\cite{kokkos-mdspan}, which also provides an implementation of the standard proposals, but to add these implementations directly in {\dunecommon}.

\subsubsection{Some additions to ranges and containers}
The data structures in {\dunecommon} are constantly being improved. For example, the \texttt{Dune::FieldVector} container, used to represent coordinates in geometries, is now mostly \texttt{constexpr}. Similarly, the container \texttt{Dune::ReservedVector}, a resizable container with static capacity similar to the \texttt{std::inplace\_vector} proposal~\cite{inplace_vector}, is now mostly \texttt{constexpr} and closely follows the \texttt{std::vector} interface.

In {\dune} we have extended the interface of iterators, especially of iterators over sparse data structures, with an additional method \texttt{.index()}, which returns the index of the currently iterated element over the dimension of the vector space. We call a range providing such an indexed iterator an ``indexed range''. With the mixin class \texttt{Dune::IndexedIterator} any iterator can be transformed into such an extended interface by simply adding an enumeration index to the iterator, which is initialized in the constructor and incremented on the iterator increment. Note that this index cannot know the position of the referenced element within a possibly larger vector space, but just numbers all the elements. An iterator of a sparse container needs to provide this index method directly with more knowledge about the position of the elements.

C++20 also introduces the very powerful \texttt{<ranges>} library. However, it is not available in older compilers and standard library implementations with which we try to remain compatible. As utilities that mimic some of its functionality, we have introduced the \texttt{Dune::TransformedRangeView} and the \texttt{Dune::TransformedRangeIterators}. These wrapper classes allow to map the original value or iterator when dereferencing the iterator of the transformed range. This gives a lot of flexibility when writing algorithms. For example, we have added the \texttt{Dune::sparseRange} utility, which wraps an indexed range to provide the index and element as a pair during traversal. An example application is iterating over the rows of a matrix, where we want to generically support dense and sparse matrices:
\renewcommand*{\thecodefile}{sparserange.cc}
\begin{cppcode}
#include <dune/common/dynvector.hh>
#include <dune/common/rangeutilities.hh>

auto A = {...}; // could be a sparse Dune::DiagonalMatrix, or a dense FieldMatrix, for example.

// simple dense/sparse matrix-vector-multiply
auto x = DynamicVector<double>(A.M(), 1.0);
auto y = DynamicVector<double>(A.N(), 0.0);
for (auto i : Dune::range(A.N()))
  for (auto&& [A_ij, j] : Dune::sparseRange(A[i]))
    y[i] += A_ij * x[j];
\end{cppcode}
A general transformed range can be created with \texttt{transformedRangeView(range, map)} or \texttt{iteratorTransformedRangeView(range, map)}, with \texttt{map} a function object taking either a reference to the element, or the iterator to the element as argument and returning the transformed element.

Matrices in {\dune} also got an upgrade. We now support lazy and real transposition of matrices, using \texttt{transposedView(matrix)} and \texttt{matrix.transposed()} respectively. The (lazy) transposed view provides the matrix-matrix multiplication and matrix-vector multiplication using \texttt{.mv} and \texttt{.mtv} and is intended for efficient operations such as $A B^T$ or $A^T B$, which are typical for geometric transformations of derivatives by the Jacobian of a geometry map, e.g.,
\renewcommand*{\thecodefile}{transposed.cc}
\begin{cppcode}
#include <dune/common/transpose.hh>
auto C = A*transposedView(B);
\end{cppcode}
The member function \texttt{.transposed()} always returns a new representation of the transposed matrix. In some cases, this may be more efficient than the view implementation. So we have added another utility that switches between the two depending on the type of the argument: \texttt{transpose(matrix)}. If \texttt{matrix} provides a \texttt{.transposed()} member function, it will be called, otherwise a view will be returned. If you pass a \texttt{std::reference\_wrapper} to a matrix, a view will always be returned. This utility is used to extend the geometry interface with (non-transposed) Jacobian and Jacobian inverse functions, see \cref{sec:extended-geometry-interface}.

The need to implement custom iterators shows up everywhere in {\dune}. In {\dunegrid} we want to iterate over the elements of a grid, in {\dunegeometry} over the sub-entities of a reference element, in {\dunecommon} in the mdspan/mdarray data structures we want to iterate over a tensor product of dimensions, etc. Writing iterators can be tedious. In the C++ standard proposal \texttt{std::iterator\_interface}~\cite{iterator_interface} introduces an interface class to help developers write iterator classes by allowing them to implement only a few necessary operations, while the rest is provided automatically. A C++17 conforming version of this proposal is implemented as a \texttt{Dune::IteratorFacade} template mixin class.

\subsubsection{Sparse matrix performance enhancement}
We introduce improvements to the sparsity pattern class \texttt{Dune::MatrixIndexSet} and sparse matrix \texttt{Dune::BCRSMatrix}. Setting up the pattern for a sparse matrix can take a significant amount of time during the assembly procedure, especially if the actual assembly can be nicely parallelized and the pattern needs to be updated frequently. The \texttt{MatrixIndexSet} collects all indices and stores them in a sorted container. For storing the element indices of a row, we switched from a classic \texttt{std::set} to an implementation similar to a \texttt{std::flat\_set}~\cite{flatset}, which uses a sorted \texttt{std::vector} internally if the number of entries is below a threshold. This has already improved the setup time of the \texttt{MatrixIndexSet}. Another improvement came from exporting the indices to the sparse matrix. Since the indices are already sorted, this could be used to speed up the setup of internal data structures in the \texttt{BCRSMatrix}. The matrix is extended by a method \texttt{.setIndicesNoSort()}, which does not perform an additional time-consuming sorting of the row indices, since this is already done. The new function \texttt{MatrixIndexSet::columnIndices(row)} returns the range over the sorted indices in the given row that can be used to fill the \texttt{BCRSMatrix}.

The performance improvement of these changes to the \texttt{MatrixIndexSet} is summarized in \Cref{tab:benchmark-matrix-indexset}. It shows a significant improvement in the setup time of the \texttt{BCRSMatrix} when assembling a Stokes equation using {\dunefunctions}.

\renewcommand*{\thecodefile}{benchmark-matrixindexset.cc}
\begin{table}[ht]
    \centering
    \begin{tabular}{cccc}
       code version & assemble pattern & \texttt{setupMatrix()} & assemble matrix \\\midrule
        old & 3.7914 sec & 2.7249 sec & 3.2119 sec \\
        new & 2.1502 sec & 0.6946 sec & 3.1353 sec \\
    \end{tabular}
    \caption{\codelink. Comparison of times for different stages of a matrix assembly process of a Taylor--Hood discretization of the Stokes equation. The first column ``assemble pattern'' measures the time for calls to \texttt{MatrixIndexSet::add(i,j)} for all indices \texttt{i,j} in the matrix pattern. The second column measures the time to create the \texttt{BCRSMatrix} and its non-zero structure, and the third column measures the time to access existing matrix entries to set values.}\label{tab:benchmark-matrix-indexset}
\end{table}

\subsubsection{Matrix pattern visualization with SVG}
\begin{figure}
    \centering
    \begin{subfigure}[b]{.32\textwidth}
        \includegraphics[width=\textwidth]{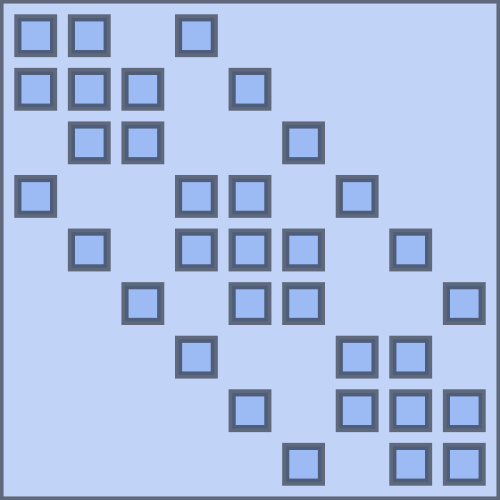}
        \caption{\texttt{T=F}}\label{fig:mat-bcrs}
    \end{subfigure}
    \hfill
    \begin{subfigure}[b]{.32\textwidth}
        \includegraphics[width=\textwidth]{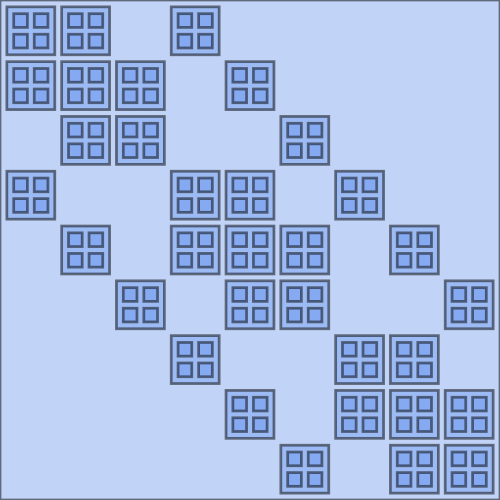}
        \caption{\texttt{T=FieldMatrix<F,2,2>}}\label{fig:mat-bcrs-fmat}
    \end{subfigure}
    \hfill
    \begin{subfigure}[b]{.32\textwidth}
        \includegraphics[width=\textwidth]{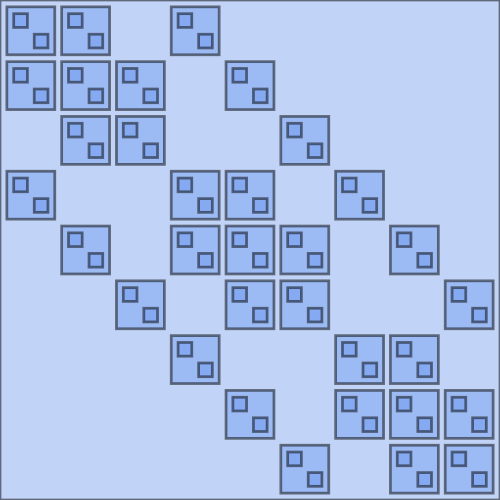}
        \caption{\texttt{T=BCRSMatrix<F>}}\label{fig:mat-bcrs-bcrs}
    \end{subfigure}
    \caption{Sample output for different nested matrix types \texttt{BCRSMatrix<T>} for field type \texttt{F}.}
    \label{fig:mat-simple}
\end{figure}

Many of the matrices in {\dune} allow for nesting of the argument types so that we can represent certain blocking structures. When more than one nesting is specified, it becomes hard to reason about these matrix patterns by just looking at the matrix indices. This is why we have implemented Scalable Vector Graphics (SVG) writer for {\dune} matrices: \texttt{Dune::writeSVGMatrix}. Figure \ref{fig:mat-simple} shows possible outputs for different nested BCRS matrices.

 \subsection{Towards a modern CMake build environment}\label{sec:buildsystem}

Beginning with version~2.4, {\dune} transitioned from the legacy Autotools-based build system to a CMake-based environment~\cite{dune:2.4}. This shift significantly streamlined module configuration, compilation, and integration workflows. The original implementation, however, adhered largely to pre-3.0 CMake conventions, relying heavily on global state and imperative configuration logic.

Starting with version~2.9, the build system has undergone a fundamental restructuring toward a modern, target-based CMake design. This modern approach improves modularity, correctness, and compatibility with external tooling. It also aligns {\dune} with contemporary best practices in C++ package management, facilitating integration with systems such as Conan~\cite{conan}, vcpkg~\cite{vcpkg}, and Spack~\cite{spack}\footnote{The implementation of an integration of {\dune} modules into C++ package managers is still work in progress.}, as well as support for CMake's \texttt{FetchContent} and superbuild mechanisms. As part of this migration, {\dune} now requires CMake version~3.16 or higher.

\subsubsection{An extended \texttt{dune\_add\_library} command}

To support this modernized build structure, each {\dune} module that defines reusable functionality is now expected to export a properly configured CMake library target. This includes specifying metadata such as namespace and export name. For this purpose, the \texttt{DuneMacros} module in {\dunecommon} provides the extended \texttt{dune\_add\_library} command, which wraps the standard \texttt{add\_library} command of CMake and enriches the target with export and packaging information.

This abstraction ensures that {\dune} libraries can be cleanly exported and reused both within the {\dune} ecosystem and by external CMake-based projects. It also reduces boilerplate in module \texttt{CMakeLists.txt} files and improves the consistency of project configurations across the framework.

\renewcommand*{\thecodefile}{}
\begin{cmakecode}
# you can create a library that gets automatically exported
dune_add_library(dunefoo EXPORT_NAME Foo NAMESPACE Dune::) # -> creates dunefoo and exports Dune::Foo
# you can add target properties to the library
target_sources(dunefoo PRIVATE foo.cc) # -> add sources to the library
target_link_libraries(dunefoo PUBLIC Dune::Common) # -> consumes Dune::Common into dunefoo
# you can consume exported target elsewhere (maybe another module)
add_executable(fooexec fooexec.cc)
target_link_libraries(fooexec PRIVATE Dune::Foo) # -> consumes Dune::Foo into fooexec
\end{cmakecode}
If the module does not compile a library, e.g., is header only or does not contain sources, the additional flag \texttt{INTERFACE} can be used in \texttt{dune\_add\_library} as in the regular \texttt{add\_library} command of CMake. The namespace and export name are used as seen in the example: in the \texttt{target\_link\_library} line, \texttt{Dune::Common} is an imported target provided by a previous call to \texttt{find\_package(dune-common)}. The namespace \texttt{Dune::} is used in all {\dune} core modules, but is not a hard requirement, e.g., discretization modules on top of {\dune} can choose their own namespace. Examples of other namespaces are \texttt{Dumux::} and \texttt{AMDiS::}. On the other hand, the \texttt{EXPORT\_NAME} argument defines the name of the component inside the namespace, typically in title case, e.g., \texttt{Common}, \texttt{Geometry}, \texttt{ALUGrid}, and typically related to the C++ namespaces used in some of the {\dune} modules. In some cases the export name corresponds to the namespace prefix itself, e.g., in {\amdis}~\cite{Praetorius2025AMDiS} the exported library would be \texttt{AMDiS::AMDiS} since there are no components to identify separately.

\subsubsection{Target properties instead of global states}
As seen in the previous example, we explicitly set properties to the library target \texttt{dunefoo} and discourage any use of global states. This includes global include directories, compile definitions, and the enforced C++ standard. In the \texttt{Dune::Common} target, which is now exported by the {\dunecommon} module, the required C++ standard is publicly enforced as a compile feature \texttt{cxx\_std\_17}. This means that binaries of this library, as well as any target that consumes it, will also transitively require C++17. If a module needs a higher C++ standard, it should firstly communicate this as a compile feature on its own module library, and secondly, since it is recommended to build the entire {\dune} module stack with the same toolchain and flags, the global CMake variable \texttt{CMAKE\_CXX\_STANDARD} should be set to the version required.

\subsubsection{Autogenerated config-headers}
A second change in the build system is the way a config header is automatically generated. Prior to {\dune} version 2.10, a \texttt{config.h} was generated based on CMake variables of all {\dune} modules the current module depends on. This \texttt{config.h} file was also required to be included in all source files. It contains preprocessor constants and macros that are defined at configure time, e.g., \texttt{\#define HAVE\_PKG 1} to indicate that a package \texttt{PKG} was found, or the version of the linked {\dune} modules, e.g., \texttt{\#define DUNE\_COMMON\_VERSION "2.10"}. Since the final \texttt{config.h} file had to be generated by the consuming module, {\dune} modules were not fully configured until the {\dune} build system helped generate it. We have changed this procedure by now generating module-local config files, e.g., \texttt{dune-foo-config.hh}, at configure time. These config files can even be included in header files of the module where the preprocessor definitions are actually needed. Thus, the consuming module does not necessarily need to generate and include a config header anymore, but just use the {\dune} headers of the modules it depends on. For backward compatibility, the old \texttt{config.h} will still be generated in a similar way as before, but will now reuse the existing module-specific config file.
 \subsection{New local finite elements}

The {\dunelocalfunctions} module provides a collection of local finite element definitions used to instantiate global function spaces across various grids. These implementations are either tailored to specific \texttt{GeometryType}s or designed generically to support arbitrary shapes built recursively via prismatic or conical extensions from reference geometries~\cite{DednerNolte2012Generic}.

In version~2.10, this collection has been further expanded. Notably, the lowest-order Raviart--Thomas elements have been extended to support both pyramid and prism geometries, following the construction presented in~\cite{Ainsworth_2017}. These additions enhance the framework's ability to support $H(\mathrm{div})$-conforming discretizations on hybrid grids that include these more complex cell types, thereby broadening the applicability of {\dune} in mixed-geometry finite element methods.
\renewcommand*{\thecodefile}{raviartthomas.cc}
\begin{cppcode}
#include <dune/localfunctions/raviartthomas/raviartthomas0prism.hh>
#include <dune/localfunctions/raviartthomas/raviartthomas0pyramid.hh>

int face_orientation_prism = 0b00001;
auto rt0_prism = Dune::RT0PrismLocalFiniteElement<double,double>{face_orientation_prism};

int face_orientation_pyramid = 0b00101;
auto rt0_pyramid = Dune::RT0PyramidLocalFiniteElement<double,double>{face_orientation_pyramid};
\end{cppcode}
In the example above, the face orientation is a bitset indicating for each face whether the normal of that face in the reference element needs to be flipped compared to the physical element in the grid.

The \texttt{Dune::HierarchicalP2WithElementBubbleLocalFiniteElement}, a hierarchical second order Lagrange finite element on simplex geometries enriched by a single bubble function in the element interior, is extended generically to arbitrary dimensions. This allows to implement stable Stokes pairs, like the conforming Crouzeix--Raviart element $(P_2+B, P_1^\text{disc})$~\cite{CR1973Conforming}.

Another addition is the hierarchical first order Lagrange finite element on simplex geometries enriched by a single bubble function, which is defined in \texttt{Dune::HierarchicalP1WithElementBubbleLocalFiniteElement}. It is an ingredient for the stable Stokes pair $(P_1+B, P_1)$, called the MINI element~\cite{ArnoldEtAl1984Stable}.
\renewcommand*{\thecodefile}{hierarchicalwithbubble.cc}
\begin{cppcode}
#include <dune/localfunctions/hierarchical/hierarchicalp1withelementbubble.hh>
#include <dune/localfunctions/hierarchical/hierarchicalp2withelementbubble.hh>

// create the P1+B finite element in 2-dimensions
auto p1b_2d = Dune::HierarchicalP1WithElementBubbleLocalFiniteElement<double,double,2>{};

// create the P2+B finite element in 3-dimensions
auto p2b_3d = Dune::HierarchicalP2WithElementBubbleLocalFiniteElement<double,double,3>{};
\end{cppcode}

\section{Development in {\dune} extensions}\label{sec:development-in-extensions}
Extension modules are modules that provide additional features and functionality to the {\dune} environment which is not covered by the core modules. In fact, most applications written with {\dune} make use of one or more extension modules. An overview of extension modules is given on the \href{https://www.dune-project.org/groups/extension/}{{\dune} website}.

\subsection{{\dunealugrid}}\label{sec:dunealugrid}
The grid \texttt{Dune::ALUGrid} provides a parallel, unstructured, conforming and non-conforming, simplex and cube, 2d and 3d grid manager. In version 2.10, the {\dunealugrid} module has seen only minor changes apart from some bug fixes.
On the Python side, the \cpp{conforming} refinement flag is now optional as a template parameter for simplex grids. When creating a \pyth{aluConformGrid} a \cpp{Dune::ALUGrid< dim, dimworld, Dune::simplex, ctype>} grid is returned and the refinement is set dynamically. This template parameter will be removed in an upcoming release. Exporting load balancing options to the Python side for grid construction now allows multiple options, including space-filling curve algorithms, the METIS~\cite{metis,Karypis2011} and ZOLTAN~\cite{zoltan,ZoltanIsorropiaOverview2012} partitioners, see \texttt{help(aluCubeGrid)} for a detailed list.

\subsection{{\dunecurvedgrid}}\label{sec:dune-curvedgrid}
The modules {\dunecurvedgeometry} and {\dunecurvedgrid} were developed to provide parametrized geometries and a grid wrapper \texttt{Dune::CurvedGrid}, see~\cite{dune-curvedgeometry,dune-curvedgrid,PraetoriusStenger2022CurvedGrid} for details. We have now moved the geometry parametrization in a modified form to the {\dunegeometry} module. The {\dunecurvedgrid} meta-grid is still based on the geometry types from {\dunecurvedgeometry}, but will adapt to the new \texttt{Dune::MappedGeometry} and \texttt{Dune::LocalFunctionGeometry} classes from {\dunegeometry} in future releases.

\subsection{{\dunefunctions}}\label{sec:dune-functions}
In {\dunefunctions} version 2.9, we already changed the interface of the global basis implementations, the pre-bases. It is no longer necessary to parameterize the pre-bases with the \texttt{MultiIndex} type returned by the \texttt{LocalView} associated with the basis. The \texttt{MultiIndex} type is defined directly inside the \texttt{LocalView} class where it is needed. It is either a \texttt{Dune::Functions::StaticMultiIndex}, essentially a \texttt{std::array} with some interface extensions, if the basis uses balanced multi-indices, i.e., the length of each multi-index is the same. Or it is a \texttt{Dune::ReservedVector}, since we always know the maximal length of a multi-index statically. Instead of specifying the multi-index type directly in the pre-bases, these classes only need to specify their maximal and minimal multi-index size, and additionally a maximal buffer size that is temporarily required during index construction.

The multi-index size information is essentially the same for most leaf-bases, which can be addressed by a single index. In addition, some member functions are also implemented identically for all leaf-bases. This has motivated us to introduce a base class \texttt{Dune::Functions::LeafPreBasisMixin} for these pre-bases, which adds common functionality to a pre-basis implementation by CRTP-like\footnote{CRTP, Curiously Recurring Template Pattern, is a technique for static polymorphism where the base class is parametrized with the type of the derived class.} inheritance:
\renewcommand*{\thecodefile}{}
\begin{cppcode}
template<class GV, int k, class R>
class LagrangePreBasis :
  public LeafPreBasisMixin< LagrangePreBasis<GV,k,R> >;
\end{cppcode}

During the implementation of several bases, we found that the code for global bases based on a local finite element with at most a single basis function associated with each sub-entity of the mesh can be easily implemented using the \texttt{Dune::MCMGMapper}. This mapper takes the layout of the basis functions, i.e., the number of functions per \texttt{Dune::GeometryType}, as a layout argument and calculates all index offsets automatically. We have added a corresponding mixin base class \texttt{Dune::Functions::LeafPreBasisMapperMixin} to facilitate basis implementations based on this approach.

A third mixin base class is the \texttt{Dune::Functions::LFEPreBasisMixin}, which goes one step further and also defines the basis node and all internal functions based on a given local finite element that implements the {\dunelocalfunctions} interface.

Some finite element bases have been added or extended in {\dunefunctions} based on the mixin classes explained above.
An example is the new \texttt{Dune::\allowbreak Functions::\allowbreak RefinedLagrangeBasis<GridView,k>}, which represents a piecewise $P_k$ Lagrange basis on a simply refined simplex element, also called $\text{iso}P_k$. An application of such a basis is the discretization of velocity components in combination with a linear Lagrange basis for the pressure in a stable Stokes pair  $(\text{iso}P_2, P_1)$~\cite{BercovierPironneau1979Error}. The basis functions of $\text{iso}P_2$ are associated with the same \texttt{GeometryType}s as those of a $P_2$ Lagrange basis.
\renewcommand*{\thecodefile}{refinedlagrangebasis.cc}
\begin{cppcode}
#include <dune/functions/functionspacebases/compositebasis.hh>
#include <dune/functions/functionspacebases/lagrangebasis.hh>
#include <dune/functions/functionspacebases/powerbasis.hh>
#include <dune/functions/functionspacebases/refinedlagrangebasis.hh>
using namespace Dune::Functions::BasisFactory;

auto stokesBasis = makeBasis(gridView,
  composite( power<2>(refinedLagrange<1>()), lagrange<1>() ) );
\end{cppcode}
We have implemented the piecewise linear ($\texttt{k=1}$) and the piecewise constant ($\texttt{k=0}$) version of this basis on a once refined simplex element.
Also, the \texttt{Dune::\allowbreak Functions::\allowbreak HierarchicalLagrangeBasis} is updated and implemented based on the \texttt{Dune::Functions::LFEPreBasisMixin}, which requires the same number of basis functions per \texttt{GeometryType} as the standard Lagrange basis. It allows to create hierarchically nested basis functions of Lagrange type up to order two, e.g.,
\renewcommand*{\thecodefile}{hierarchicallagrangebasis.cc}
\begin{cppcode}
#include <dune/functions/functionspacebases/hierarchicallagrangebasis.hh>
using namespace Dune::Functions::BasisFactory;

auto hP2Basis = makeBasis(gridView, hierarchicalLagrange<2>());
\end{cppcode}
Note that this hierarchic form of a Lagrange basis can also be used to define a stable Stokes finite element.

A meta basis similar to the existing \texttt{Dune::\allowbreak Functions::PowerBasis} is introduced to allow for specification of the number of components at run-time: \texttt{Dune::\allowbreak Functions::\allowbreak DynamicPowerBasis}. This is based on the \texttt{Dune::\allowbreak TypeTree::\allowbreak DynamicPowerNode} introduced in {\dunetypetree}. A typical application of such a basis is to represent a variable number of reactive agents in a reaction-diffusion system, for example. Thereby, the number of different agents can be something that is specified in a parameter file alongside with the different reaction coefficients. In this way, the whole system can be implemented as one code and then dynamically setup the configuration loaded from a file, see \href{https://dune-copasi.netlify.app/tutorials}{\dunecopasi}~\cite{Ospina2024DuneCopasi}.
Another application is in multiphase field problems, e.g., for the simulation of a collection of biological cells,~\cite{HappelVoigt2024Coordinated,PraetoriusVoigt2018Collective}, where each phase field represents a single cell and is coupled to the others with an interaction term. The number of cells can be chosen dynamically, given an initial shape and position in a parameter file.
\renewcommand*{\thecodefile}{dynamicpowerbasis.cc}
\begin{cppcode}
#include <dune/functions/functionspacebases/dynamicpowerbasis.hh>
using Dune::Functions::BasisFactory;

std::size_t n = /* number of cells */;
auto basis = makeBasis(gridView, power(lagrange<2>(), n));
\end{cppcode}

\subsection{{\dunevtk}}\label{sec:dune-vtk}
A file reader and writer for the VTK file format, adapted to the {\dune} grid interface and {\dunefunctions} grid-functions definitions, is added to the set of extension modules as {\dunevtk}~\cite{dune-vtk}. It supports several VTK grid formats, including unstructured grids, structured grids such as rectilinear grids, and image data. It also provides ASCII and BINARY output formats with and without compression, as well as conforming, non-conforming, quadratic and Lagrangian parametrized elements, and continuous and discontinuous data. The file reader \texttt{Dune::Vtk::VtkReader} allows to read all file types that the writer \texttt{Dune::Vtk::VtkWriter} can write.
In {\dunevtk} 2.10 all functions and classes have been moved to the \texttt{Dune::Vtk} namespace. The old classes are still available, but deprecated. We have also developed Python bindings to support reading VTK files from Python using the same interface.

\section{Further noteworthy changes}\label{sec:further-noteworthy-changes}
Several changes introduced in the {\dune} core modules not described in detail before are listed in the following section. We sort these updates by the modules.

\subsection{Infrastructure}
\begin{itemize}
\item A call to \texttt{find\_package(dune-<module>)} now resolves all direct {\dune} dependencies inside of the find module and not the full tree of dependencies.
\item The version checking macro \texttt{DUNE\_VERSION\_NEWER} is replaced by \texttt{DUNE\_VERSION\_GTE}.
  \item The CMake module \texttt{FindSuiteSparse} and the utility \texttt{add\_dune\_suitesparse\_flags} is moved from {\duneistl} to {\dunecommon}.
  \item The config variable \texttt{HAVE\_UMFPACK} is deprecated and should be replaced by \texttt{HAVE\_SUITESPARSE\_UMFPACK}.
  \item MPI related flags are moved out of the \texttt{config.h} file into compiler flags that can be set by \texttt{add\_dune\_mpi\_flags(target)} to a target.
  \item Python bindings are now enabled by default. Packages are automatically installed either in an internal virtual environment or in an active environment during the module build process.
\end{itemize}

\subsection{\dunecommon}
\begin{itemize}
\item The utility \texttt{Dune::switchCases} is extended to support both, compile-time switch and runtime switch, depending on the range type of cases.
\item The \texttt{MPIHelper} singleton is now thread-safe by default and allows creation without any initialization arguments after this is given for the first time.
  \item The utilities \texttt{CollectiveCommunication} is renamed into \texttt{Communication}.
\item The class template \texttt{Dune::SizeOf} was provide to avoid compiler bug in the implementation of the C++ language feature \texttt{sizeof...}. Since this is no longer an issue with the supported compilers it is deprecated and will be removed in before the next release of {\dunecommon}.
  \item The \texttt{DUNE\_ASSERT\_AND\_RETURN} macro is a workaround for the pre-C++14 limitation, that a classical \texttt{assert()} was not allowed inside \texttt{constexpr} functions. This macros is deprecated and will be removed in before the next release of {\dunecommon}.
  \item The utilities \texttt{Dune::Power} and \texttt{Dune::StaticPower} are deprecated in {\dune} 2.9 and are removed in {\dune} 2.10 alongside with the file \texttt{power.hh}. The better replacement for these utilities is the generic \texttt{constexpr} function \texttt{Dune::power} from \texttt{math.hh}.
\end{itemize}

\subsection{\dunegeometry}
\begin{itemize}
\item Quadrature rules are now structured-bindings aware and can iterate over the pairs $(x,w)$, position and weight, directly.
\end{itemize}

\subsection{\dunelocalfunctions}
\begin{itemize}
  \item We fixed the definition of local keys in the Crouzeix--Raviart local finite element by attaching basis functions to codim-1 entities (facets) instead of dim-1 entities (edges).
  \item The utility \texttt{PolynomialBasis} is fixed to support derivatives in 3d.
\end{itemize}

\subsection{\duneistl}
\begin{itemize}
\item The \texttt{Dune::ScaledIdentityMatrix} is extended by an \texttt{operator*} for multiplication with scalar factors.
  \item The direct solver CHOLMOD from the SuiteSparse collection now supports the integer type \texttt{long int} and allows to export its internally created factorization object.
  \item The SPQR solver for sparse QR-factorization is cleaned up to support block sparse-matrices as input and to actually allow least-squares problems by supporting non-square matrices.
  \item We made AMGs \texttt{ParallelIndicesCoarsener} more deterministic by fixing the order in which processes perform the coarsening, if the new flags \texttt{useFixedOrder} is provided.
  \item The \texttt{MinRes} iterative solver now computes the initial residual as a preconditioned residual.
  \item We have added a new preconditioner to {\duneistl}: \texttt{D-ILU}~\cite{Pommerell1992Solution} an incomplete factorization preconditioner.
\end{itemize}

\subsection{\dunegrid}
\begin{itemize}
  \item The capability \texttt{Dune::\allowbreak Capabilities::\allowbreak viewThreadSafe} is specialized for \texttt{Dune::\allowbreak YaspGrid}.
  \item A runtime capability \texttt{.isConforming()} is added to the grid interface to indicate whether an instance of the grid is conforming, even if it cannot be guaranteed as a static compile-time property.
  \item The methods \texttt{.ibegin()} and \texttt{.iend()} are moved from \texttt{Dune::\allowbreak DefaultGridView} into the class \texttt{Dune::\allowbreak GridDefaultImplementation}.
  \item The deprecated load balancing utility \texttt{YLoadBalance} from \texttt{Dune::YaspGrid} is removed. Use the new \texttt{Dune::Yasp::Partitioning} class instead.
  \item The grid views of \texttt{Dune::UGGrid} are now thread-safe which allows for thread-parallel assembly.
\end{itemize}

\subsection{\dunefunctions}
\begin{itemize}
  \item A data structure representing the shape of the index-space is exported by the pre-bases via the \texttt{.containerDescriptor()} method. It is similar to a nested tree of standard containers like array, vector, and tuple, with some special types for nodes with identical child tree-shapes.
  \item The \texttt{Dune::\allowbreak Functions::\allowbreak interpolate()} utility can now be used with
  bases where the local interpolation involves the evaluation of derivatives, e.g.
  for Hermite-type elements.
  \item The utility \texttt{DefaultNodeToRangeMap} is deprecated and should be replaced by \texttt{Dune::\allowbreak Functions::\allowbreak HierarchicNodeToRangeMap}.
  \item The namespace \texttt{Dune::\allowbreak Functions::\allowbreak BasisBuilder} is deprecated and should be replaced by \texttt{Dune::\allowbreak Functions::\allowbreak BasisFactory}.
  \item The wrapper \texttt{HierarchicalVectorWrapper}, which adds multi-index access to data structures, is deprecated and replaced by \texttt{Dune::\allowbreak Functions::\allowbreak ISTLVectorBacked}.
  \item The Nédélec and Raviart--Thomas function space bases are now accessible via the Python interface.
  \item The Python bindings of \texttt{interpolate()} can now be used with vector valued functions.
\end{itemize}

\section*{Statements and Declarations}

\subsection*{Acknowledgements}
We would further like to thank Nils-Arne Dreier,
Henrik Stolzmann,
Lukas Renelt,
Patrick Jaap,
Kilian Weishaupt,
Felix Gruber,
Lasse Hinrichsen,
Maik Porrmann,
Tobias Leibner,
Lisa Julia Nebel,
Alexander Müller,
Jakob Torben,
Eduardo Bueno,
andrthu,
Hendrik Kleikamp,
Mathis Kelm,
Max Kahnt, and
Sanchi Vaishnavi

\subsection*{How to cite DUNE}
If using one of the {\dune} core modules please cite the appropriate papers from the list of original {\dune} papers (\cite{dunepaperII:08,dunepaperI:08,BlattBastian2007DuneIstl,BlattBastian2008Generic}) and the current release notes paper. Please note that other {\dune} modules might require citation of further papers, such as {\dunealugrid} (\cite{AlkamperEtAl2016DuneALUGrid}), {\dunefem} (\cite{DednerEtAl2010DuneFem}), {\dunepdelab} (\cite{BlattHeimannMarnach2010Generic}), or {\dunecurvedgrid} (\cite{PraetoriusStenger2022CurvedGrid}). Further {\dune} modules are described in~\cite{BastianEtAl2021Dune} and on the {\dune} web page \href{https://www.dune-project.org}{dune-project.org}.

\subsection*{Funding}
C.\ Engwer was partially supported by the German Research Foundation (DFG) through projects HyperCut I \& II (project number 439956613), and EsCut (project number 526031774) within the DFG Priority Research Program 2410 CoScaRa ``Hyperbolic Balance Laws in Fluid Mechanics: Complexity, Scales, Randomness''. C.\ Engwer further acknowledges support by the German Research Foundation (DFG) under Germany's Excellence Strategy EXC 2044-390685587, Mathematics M\"{u}nster: Dynamics--Geometry--Structure.

S.\ Ospina De Los R{\'\i}os was supported by the German Federal Ministry of Education and Research (BMBF) through the project FKZ 031L0158 ``STML-Tools: Tools for cell biological spatio-temporal models in life sciences''.

S.\ Praetorius was partially supported by the German Research Foundation (DFG) through the research unit FOR3013, ``Vector- and Tensor-Valued Surface PDEs'', within the project TP06, ``Symmetry, length, and tangential constraints'' (project number 417223351).

\subsection*{Conflict of interest}
The authors have no competing interests to declare that are relevant to the content of this article.

\bibliography{references.bib}

\clearpage
\begin{appendices}

\section{Supplemental material: The full source code examples}

\subsection{concepts.cc}\label{code:concepts.cc}
\inputcpp{examples/concepts.cc}

\subsection{gridconcepts.cc}\label{code:gridconcepts.cc}
\inputcpp{examples/gridconcepts.cc}

\subsection{indexrange.cc}\label{code:indexrange.cc}
\inputcpp{examples/indexrange.cc}

\subsection{hybridfunctor.cc}\label{code:hybridfunctor.cc}
\inputcpp{examples/hybridfunctor.cc}

\subsection{integersequence.cc}\label{code:integersequence.cc}
\inputcpp{examples/integersequence.cc}

\subsection{mappedgeometry.cc}\label{code:mappedgeometry.cc}
\inputcpp{examples/mappedgeometry.cc}

\subsection{localfiniteelementgeometry.cc}\label{code:localfiniteelementgeometry.cc}
\inputcpp{examples/localfiniteelementgeometry.cc}

\subsection{mdspan.cc}\label{code:mdspan.cc}
\inputcpp{examples/mdspan.cc}

\subsection{sparserange.cc}\label{code:sparserange.cc}
\inputcpp{examples/sparserange.cc}

\subsection{transposed.cc}\label{code:transposed.cc}
\inputcpp{examples/transposed.cc}

\subsection{benchmark-matrixindexset.cc}\label{code:benchmark-matrixindexset.cc}
\inputcpp{examples/benchmark-matrixindexset.cc}

\subsection{raviartthomas.cc}\label{code:raviartthomas.cc}
\inputcpp{examples/raviartthomas.cc}

\subsection{hierarchicalwithbubble.cc}\label{code:hierarchicalwithbubble.cc}
\inputcpp{examples/hierarchicalwithbubble.cc}

\subsection{refinedlagrangebasis.cc}\label{code:refinedlagrangebasis.cc}
\inputcpp{examples/refinedlagrangebasis.cc}

\subsection{hierarchicallagrangebasis.cc}\label{code:hierarchicallagrangebasis.cc}
\inputcpp{examples/hierarchicallagrangebasis.cc}

\subsection{dynamicpowerbasis.cc}\label{code:dynamicpowerbasis.cc}
\inputcpp{examples/dynamicpowerbasis.cc}
 \end{appendices}

\end{document}